\documentclass[10pt,aps,prd,twocolumn,showpacs,superscriptaddress,eqsecnumnofootinbib,longbibliography,nofootinbib]{revtex4-1}
\usepackage{mathrsfs,amsmath,amsthm,latexsym,amssymb,amsfonts,epsfig,txfonts,cancel}
\usepackage{xcolor}
\definecolor{navy}{RGB}{0,0,150}
\usepackage[colorlinks, linkcolor=cyan, citecolor=magenta, urlcolor=navy, plainpages=false, pdfstartview=FitH]{hyperref}
\usepackage{appendix}
\allowdisplaybreaks
\newcommand{\GZU}{School of Physics, Guizhou University, Guiyang 550025, China}
\newcommand{\BNU}{Department of Physics, Beijing Normal University, Beijing 100875, China}
\newcommand{\st}{\scriptstyle}

\begin{document}

\title{Light bending and gravitational lensing in Brans-Dicke theory}

\author{Xiaojun Gao}
\affiliation{\GZU}

\author{Shupeng Song}
\email{songsp@mail.bnu.edu.cn}
\affiliation{\BNU}

\author{Jinsong Yang}
\thanks{Corresponding author}
\email{jsyang@gzu.edu.cn}
\affiliation{\GZU}


\begin{abstract}
As an important candidate theory of gravity, Brans-Dicke theory has been widely studied. In this paper, we investigate light bending and gravitational lensing by compact objects in Brans-Dicke theory in weak gravitational field. Firstly, we present a general formalism for calculating higher-order corrections to light bending angle and lensing observables for a static, spherically symmetric and flat spacetime, in which the metric is given in the isotropic coordinates. Secondly, we apply the general formalism to Brans-Dicke theory and get the corresponding light bending angle and lensing observables. Our results show that, although the sums over the low-order correction terms in magnifications of the primary and secondary images do not dependent on the theories of gravity, the sums over correction terms with order higher than three do. Moreover, we show that the total magnification has a non-vanishing first-order correction, rather than a vanishing contribution concluded in the literature. We find that the corrections to lensing observables of BD theory close to those of GR when the parameter $\omega$ tends to $+\infty$ from $-\frac32$, while opposition occurs when $\omega$ tends to $-2$ from $-\infty$.
\end{abstract}


\maketitle
\newpage

\section{Introduction}

It is well know that Einstein's general relativity (GR) providing gravity with a geometric description has been confirmed in all observations and experiments up to now. Other alternative theories of gravity were also proposed under different considerations. Brans and Dicke proposed in 1961 a very competitive gravitational theory -- Brans-Dicke theory of gravity (BD theory) \cite{Brans:1961sx}. BD theory belongs to a sort of the simplest and most important scalar-tensor gravitational theory. In BD theory, the gravitational field is determined by both a scalar field $\phi$ and a metric tensor field $g_{ab}$. Compared with GR, BD theory can not only pass through all gravitational experiments to date, but also explain naturally the accelerated expansion of the universe without introducing the dark energy \cite{Will:2014kxa,Bertolami:1999dp,Qiang:2004gg,Bisabr:2011re}. BD theory recovers to GR as parameter $\omega$ goes to infinity. Therefore, BD theory is regarded as a natural generalization of GR.

To distinguish different theories of gravity, some suggestions based on gravitational effects are proposed. As an important effect of gravitation, the light bending by the sun was predicted by Einstein in 1916 \cite{Einstein:1916vd}. The bending angle predicted by GR has been confirmed to a high accuracy by many observations since 1919 \cite{Dyson:1920cwa,Lebach:1995zz}. The observable phenomena resulting from light bending is called as gravitational lensing (GL), and the corresponding gravitational source is known as a gravitational lens. GL has been studied with great interest in astrophysics as well as in theoretical physics \cite{Einstein:1956zz,Zwicky:1937zzb,Schneider:1992bk,Wambsganss:1998gg,Perlick:2004tq,Bozza:2009yw}. The first example of GL was discovered in 1979 \cite{Walsh:1979nx}. According to the strength of gravitational field, GL is divided into strong GL \cite{Dey:2012ph,Bozza:2001xd,Eiroa:2002mk,Bozza:2005tg,Bozza:2006nm} and weak GL \cite{Sereno:2003nd,Keeton:2005jd}. With the help of GL, one can not only obtain some informations about the distant and dim stars, but also probe the dark matter, dark energy and the properties of some strange substances in the universe. Furthermore, GL is widely used to test and distinguish different theories of gravity \cite{Keeton:2005jd,Keeton:2006sa}. It is also regarded as one of the most important tools for quantifying the mass content and distribution in distant galaxies, and for distinguishing naked singularities from black holes \cite{Virbhadra:2002ju,Virbhadra:2007kw,Gyulchev:2008ff,Sahu:2012er}. Hence light bending and GL have been active subjects in theoretical research as well as in astronomical observation. 

Given a static, spherically symmetric metric, we can write it in two convenient coordinate systems, namely the Schwarzschild (standard) and the isotropic coordinate systems. In principle, one can compute in one of coordinates corrections to light bending angle and lensing observables, expressed in terms of the conserved impact parameter $b$, and obtain the same coordinate-independent results. However, in practice, the metric describing a spacetime has a rather compact formalism in one coordinates than the others. Therefore, let alone an exact analytical relation, even an approximation relation between these two coordinate systems, in their parameterization-post-Newtonian (PPN) formalisms up to a certain order, can not be built. A general formalism of corrections to light bending angle and lensing observables in corresponding orders obtained in one coordinate system can not be conveniently transformed to the other coordinate system. The formalism of these corrections obtained in the Schwarzschild coordinates has been presented in \cite{Keeton:2005jd,Keeton:2006sa}.

In this paper, we study light bending and GL in BD theory. To that end, we first develop a general formalism to calculate corrections of light bending angle to fourth order as well as of lensing observables to third order for a static, spherically symmetric and asymptotically flat metric expressed in the isotropic coordinates. We then apply the general formalism to BD theory in order to explore the difference between BD theory and GR in the lensing observables.

The rest of this paper is organized as follows. In Sec. \ref{sec:II}, in the isotropic coordinate system, we present a general integration expression of light bending angle in a static, spherically symmetric spacetime around a compact object. Then the PPN metrics are used to calculate the correction to light bending angle as a function of the invariant impact parameter up to fourth order, as well as corrections to observable properties of lensed images (positions, magnifications) up to third order. In Sec. \ref{sec:III}, applying the general formalism derived in Sec. \ref{sec:II} to BD theory, we obtain the actual form of light bending angle, positions and magnifications of the lensed images. In Sec. \ref{sec:IV}, we summarize and discuss our results. Throughout this paper, we use the geometrized units of $G=c=1$.

\section{Light bending and GL in the PPN formalism}
\label{sec:II}

In this section, for a static, spherically symmetric and asymptotically flat metric expanded as PPN formalism in the isotropic coordinates, we will derive a general formalism for computing corrections to light bending angle and lensing observables in terms of the invariant impact parameter.

\subsection{Bending angle}

For a static, spherically symmetric and asymptotically flat spacetime, its line element can be written in the isotropic coordinates $(t,\rho,\theta,\varphi)$ as
\begin{align}\label{eq:1}
 {\rm d}s^{2}=-A(\rho){\rm d}t^2+B(\rho)\left[{\rm d}\rho^2+\rho^2\left({\rm d}\theta^2+\sin^2\theta\,{\rm d}\varphi^2\right)\right]\,,
\end{align}
where $A(\rho)\rightarrow1$ and $B(\rho)\rightarrow1$ as $\rho\rightarrow\infty$.

Consider a light ray coming from a source in the flat region along a null geodesic  is deflected by a compact object (lens), and then arrives at an observer in the flat region. See Fig. \ref{lensing} for the schematic diagram. Without loss of generality, the geodesics of the light ray can be taken on the equatorial plane ($\theta=\pi/2)$. Then the orbital equation of the light ray can be written as \cite{Weinberg:1972kfs,Bozza:2009yw}
\begin{align}\label{eq:2}
 \frac{{\rm d}\varphi}{{\rm d}\rho}=\frac{1}{\rho^2}\sqrt{\frac{A(\rho)}{\frac{B(\rho)}{b^2}-\frac{A(\rho)}{\rho^2}}}\,,
\end{align}
where $b\equiv\frac{L}{E}$ is the so-called impact parameter, here $E$ and $L$ are the total conserved energy and angular momentum at infinity, respectively. Denote the light ray's radial distance of closet approach to the lens by $\rho_0$ at $\varphi=0$. Consider $\left.{\rm d}\rho/{\rm d}\varphi\right|_{\rho=\rho_0}=0$, the impact parameter $b$ is related to $\rho_0$ by
\begin{align}\label{eq:4}
 b=\left[\frac{B(\rho_0)\rho_0^2}{A(\rho_0)}\right]^{1/2}\,.
\end{align}
The light bending angle can be obtained from Eq. \eqref{eq:2}, which is given by
\begin{align}\label{eq:3}
 \hat{\alpha }(\rho_0)&=2\int^{\infty}_{\rho_0} \frac{{\rm d}\rho}{\rho^2}\left[\frac{A(\rho)}{\frac{B(\rho)}{b^2}-\frac{A(\rho)}{\rho^2}}\right]^{1/2}-\pi\,.
\end{align}
It is convenient to express the integral in Eq. \eqref{eq:3} in terms of a new variable $x=\rho_0/\rho$ as
\begin{align}\label{eq:5}
 \hat{\alpha}(\rho_0)&=2\int^{1}_{0}{\rm d}x\left[\frac{A(\rho_0/x)}{B(\rho_0/x)\left(\frac{\rho_0}{b}\right)^2-A(\rho_0/x)\,x^2}\right]^{1/2}-\pi\,.
\end{align}
Hence the bending angle $\hat{\alpha }(\rho_0)$ is determined only by $A(\rho)$ and $B(\rho)$. However, the bending angle $\hat{\alpha }(\rho_0)$ in Eq. \eqref{eq:3} or \eqref{eq:5} as elliptic integral can not be analytical evaluated particularly. In what follows, we assume that the gravitational field outside of lens is so weak that the components $A(\rho)$ and $B(\rho)$ of the spacetime metric can be expanded as a series of the small parameter $M/\rho$ with $M$ being mass of a compact object. In that case, the integral can be approximatively evaluated term by term. It turns out that knowledge of light bending angle to order $M/\rho_0$ requires knowledge of every term in the metric to the same order. 

In the following, we derive the light bending angle up to the fourth order. To do that, we expand the coefficients $A(\rho)$ and $B(\rho)$ in Eq. \eqref{eq:1} as a Taylor series to fourth order in $M/\rho$ as follows
\begin{align}\label{eq:6}
 A(\rho)=&1-2\alpha \frac{M}{\rho}+2\lambda \left(\frac{M}{\rho}\right)^2-\frac{3}{2}\xi \left(\frac{M}{\rho}\right)^3 \notag \\
 &+\kappa \left(\frac{M}{\rho}\right)^4+O\left[\left(\frac{M}{\rho}\right)^5\right],\\
 B(\rho)=&1+2\gamma \frac{M}{\rho}+\frac{3}{2}\delta \left(\frac{M}{\rho}\right)^2+\frac{1}{2}\eta \left(\frac{M}{\rho}\right)^3 \notag \\
 &+\frac{1}{16}\nu \left(\frac{M}{\rho}\right)^4+O\left[\left(\frac{M}{\rho}\right)^5\right]\,,\label{eq:7}
\end{align}
where the parameters $\alpha,\lambda,\cdots$ are PPN parameters. The PPN parameters corresponding to the Schwarzschild metric take
\begin{align}\label{eq:8}
 \alpha=\lambda=\xi=\kappa=\gamma=\delta=\eta=\nu=1\,.
\end{align}
Now we expand the term on the right hand side of  Eq. \eqref{eq:4} as a series in $M/\rho_0$
\begin{align}\label{eq:9}
 b=& \rho_0\left\{1+a_{1}\left(\frac{M}{\rho_0}\right)+a_{2}\left(\frac{M}{\rho_0}\right)^2 +
  a_{3}\left(\frac{M}{\rho_0}\right)^3 \right. \notag\\
  &\qquad\left.+a_{4}\left(\frac{M}{\rho_0}\right)^4+O\left[\left(\frac{M}
  {\rho_0}\right)^5\right]\right\}\,,
\end{align}
where the factors $a_{1},a_{2},a_{3},a_{4}$ are respectively
\begin{align}\label{eq:10}
 a_{1}=& \alpha+\gamma\,,\qquad a_{2}=\frac{1}{4}\left(6\alpha^{2}-4\lambda+4\alpha\gamma-2\gamma^{2}+3\delta\right)\,, \notag\\
 a_{3}=& \frac{1}{4}\left[10\alpha^{3}+(6\alpha^{2}-4\lambda+2\gamma^{2}-3\delta)\gamma-2\alpha(6\lambda+\gamma^{2})\right. \notag\\
       & +3\alpha\delta+3\xi+\eta \Big]\,, \notag\\
 a_{4}=& \frac{1}{32}\left[140\alpha^{4}+48\lambda^{2}+80\alpha^{3}\gamma-20\gamma^{4}+8\lambda(2\gamma^{2}-3\delta)\right. \notag\\
       &-12\alpha^{2}(20\lambda+2\gamma^{2}-3\delta)+36\gamma^{2}\delta-9\delta^{2}+24\gamma\xi-8\gamma\eta \notag\\
       &+8\alpha\left(-12\lambda\gamma+2\gamma^{3}-3\gamma\delta+9\xi+\eta\right)-16\kappa+\nu \Big]\,.
\end{align}
Inserting Eqs. \eqref{eq:6}, \eqref{eq:7} and \eqref{eq:9} into Eq. \eqref{eq:3} (or Eq. \eqref{eq:5}) and then expanding the integrated function as a series in $M/\rho_0$ to the same order, we can carry out the integration in Eq. \eqref{eq:5} term by term to obtain the bending angle
\begin{align}\label{eq:11}
\hat{\alpha }(\rho_0)=&2(\alpha +\gamma)\frac{M}{\rho_0}+\frac{1}{4}\left[8 (\pi -1) \alpha^2+8 (\pi -2) \alpha  \gamma \right.\notag\\
&-4 \pi  \lambda -8 \gamma^2+3 \pi  \delta \Big]\left(\frac{M}{\rho_0}\right)^2+\frac{e_{1}}{6}\left(\frac{M}{\rho_0}\right)^3 \notag\\
&+\frac{e_{2}}{64}\left(\frac{M}{\rho_0}\right)^4+O\left[\left(\frac{M}{\rho_0}\right)^5\right]\,,
\end{align}
where
\begin{align}\label{eq:12}
e_{1}=&(134-24\pi)\alpha^3+6(31-8\pi)\alpha^2\gamma+3\alpha\left[4(\pi-9)\lambda\right.\notag\\
      &+(22-8\pi)\gamma^2-3(\pi-5)\delta]+12(\pi-5)\lambda\gamma+14\gamma^3\notag\\
      &-9(\pi-1)\gamma\delta+6(3\xi+\eta )\,,\notag\\
e_{2}=&64(30\pi-71)\alpha^4+512(7\pi-20)\alpha^3\gamma-32\alpha^2\notag\\
      &\times\left[4(16\pi-31)\lambda+(220-68\pi)\gamma^2+3(17-7\pi)\delta\right] \notag\\
      &-16\alpha\left[48(3\pi-8)\lambda\gamma-32(\pi-3)\gamma^3+12(10-3\pi)\right. \notag\\
      &\times\gamma\delta-27\pi\xi+42\xi-6\pi\eta+14\eta\Big]+\pi\left[160\lambda^2-32\lambda\right. \notag\\
      &\times\left(20\gamma^2+3\delta\right)+3\left(64\gamma^2\delta+16\gamma(6\xi+\eta)-6\delta^2-16\right. \notag\\
      &\times\kappa+\nu\Big)\Big]-32\gamma\left[-68\lambda\gamma+6\gamma^3+9\gamma\delta+7(3\xi+\eta)\right]\,.
\end{align}
It is easy to see that the above expression $\hat{\alpha }(\rho_0)$ of bending angle in terms of the radial coordinate distance $\rho_0$ is coordinate dependent. In the following, we will derive a coordinate-independent expression for the bending angle in terms of the impact parameter $b$. 

We assume that $\rho_0$ can be expanded as a series of $M/b$
 \begin{align}\label{eq:13}
 \rho_0=& b\left\{1+c_{1}\left(\frac{M}{b}\right)+c_{2}\left(\frac{M}{b}\right)^2 +
  c_{3}\left(\frac{M}{b}\right)^3+c_{4}\left(\frac{M}{b}\right)^4\right.\notag\\
  &\hspace{3cm}\left.+O\left[\left(\frac{M}
  {b}\right)^5 \right]\right\}\,.
 \end{align}
Putting Eq. \eqref{eq:13} into Eq. \eqref{eq:9}, the factors $c_{i}$ can be determined by requiring the coefficient of each term of $(M/b)^{i}$ is equal to zero, which are given by
\begin{align}\label{eq:14}
 c_{1}=&-(\alpha+\gamma)\,,\qquad\quad c_{2}=\lambda-\frac{3\alpha^{2}}{2}-\alpha\gamma+\frac{\gamma^{2}}{2}-\frac{3\delta}{4}\,, \notag\\
 c_{3}=& 4\alpha\lambda-4\alpha^{3}+(2\lambda-4\alpha^{2})\gamma-\frac{3\alpha\delta}{2}-\frac{3\xi}{4}-\frac{\eta}{4}\,, \notag\\
 c_{4}=& \frac{1}{32}\left\{-420\alpha^{4}-80\lambda^{2}-560\alpha^{3}\gamma+48\lambda\gamma^{2}-4\gamma^{4}+20\alpha^{2} \right.\notag\\
      &\times\left(28\lambda-6\gamma^{2}-9\delta\right)+72\lambda\delta+12\gamma^{2}\delta-9\delta^{2}-72\gamma\xi-8\gamma\eta \notag\\
      &+8\alpha\left[60\lambda\gamma+2\gamma^{3}-9\gamma\delta-3(5\xi+\eta)\right]+16\kappa-\nu\Big\}\,.
\end{align}
Inserting Eq. \eqref{eq:13} into Eq. \eqref{eq:11} yields
\begin{align}\label{eq:15}
\hat{\alpha}(b)=&A_{1}\frac{M}{b}+A_{2}\left(\frac{M}{b}\right)^2+A_{3}\left(\frac{M}{b}\right)^3+A_{4}\left(\frac{M}{b}\right)^4+O\left[\left(\frac{M}{b}\right)^5\right]\,,
\end{align}
where
\begin{align}\label{eq:16}
A_1=&2(\alpha+\gamma)\,,\qquad A_2=\frac{\pi}{4}[8\alpha(\alpha+\gamma)-4\lambda+3\delta]\,,\notag\\
A_3=&\frac{70\alpha^3}{3}+30 \alpha^2\gamma+\alpha(-20\lambda+6\gamma^2+9\delta)-12\lambda\gamma \notag\\
      &-\frac{2\gamma^3}{3}+3\gamma\delta+3\xi+\eta\,, \notag\\
A_4=&\frac{3\pi}{64}\left\{2\Big[320\alpha^4+512\alpha^3\gamma-48 \alpha^2\left(8\lambda-4\gamma^2-3\delta\right)\right.\notag\\
      &+8\alpha[12\gamma(\delta-4\lambda)+9\xi+2\eta]+48\lambda^2-16\lambda \notag\\
      &\times(4\gamma^2+3\delta)+8\gamma(6\xi+\eta)+9\delta^2-8\kappa\Big]+\nu\Big\}\,.
\end{align}

\subsection{Positions and Magnifications of lensing images}

\begin{figure}[htbp]
\centering
\includegraphics[height=0.3\textwidth]{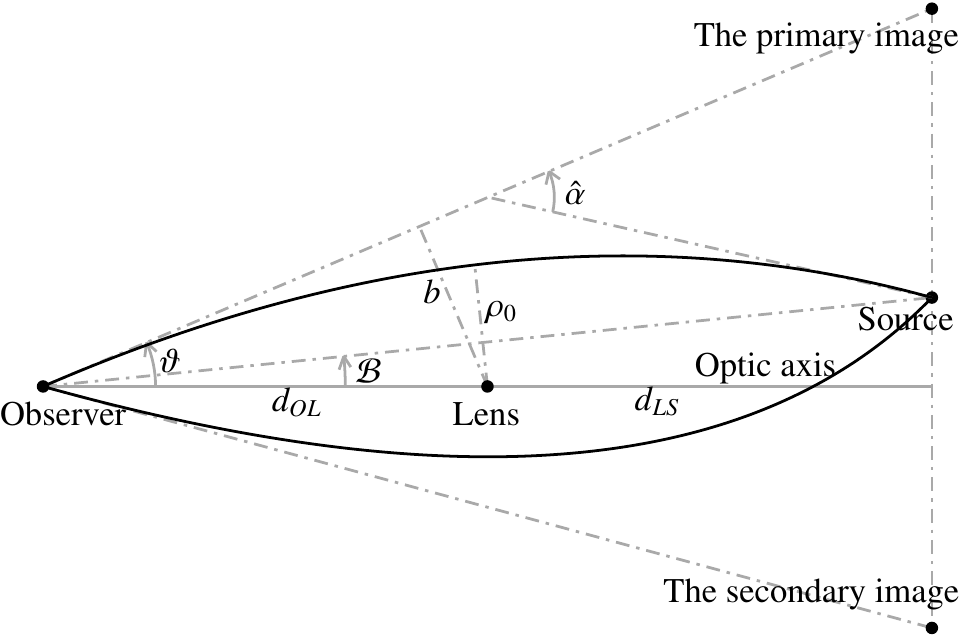}\\
\caption{The schematic diagram of light bending and GL.}
\label{lensing}
\end{figure}

The phenomenon of GL is closely related to the light bending. The geometrical picture of GL is shown in Fig. \ref{lensing}. The line connecting the observer to the lens is called as the optic axis. When the source, lens and observer are misaligned, two images arise on both sides of the optic axis. The image on the same/opposite side as/to the source with respect to the optic axis is called as the primary/secondary image. In the special case where the lens components are aligned, the image of the source is the Einstein ring around the optic axis. The gravitational lens equation is given by
\begin{align}\label{eq:17}
\tan {\cal B}=\tan\vartheta-D[\tan\vartheta+\tan(\hat{\alpha}-\vartheta)]\,,
\end{align}
where ${\cal B}$ and $\vartheta$ are the angular positions of the unlensed source and its image, respectively, $\hat{\alpha}$ is the bending angle, and $D\equiv d_{LS}/(d_{OL}+d_{LS})$, here $d_{OL},d_{LS}$ are respectively the observer-lens, and lens-source angular diameter distances. The value of $\vartheta$ will be obtained by solving Eq. \eqref{eq:17} by perturbation method in the following. Following the reference \cite{Keeton:2005jd}, we introduce a small parameter
\begin{align}
 \varepsilon:=\frac{\vartheta_\bullet}{\vartheta_{E}}\cong\frac{\vartheta_{E}}{4D}\,,
\end{align}
where $\vartheta_\bullet\equiv\tan^{-1}(M/d_{OL})$ is the angle subtended by the gravitational radius, and $\vartheta_{E}\equiv\sqrt{4MD/d_{OL}}$ denotes the angular radius of Einstein rings. The variable $\varepsilon$ $(\ll1)$ will be used as the expanding parameter. The variables ${\cal B}$ and ${\cal \vartheta}$ in Eq. \eqref{eq:17} can be regarded as functions of $\varepsilon$
\begin{align}\label{eq:variables_function}
 {\cal B}(\varepsilon)=4D\beta\varepsilon\,,\qquad {\cal \vartheta}(\varepsilon)=4D\theta\,\varepsilon\,.
\end{align}
Now, let us relate the bending angle $\hat{\alpha}$ in Eq. \eqref{eq:15}, expressed as a series in the small parameter $M/b$, to the parameter $\varepsilon$. To that end, we need to build a relation between these two small parameters $M/b$ and $\varepsilon$. Considering the geometric relations $\sin\vartheta=b/d_{OL}$ and $\tan\vartheta_\bullet=M/d_{OL}$, we have
\begin{align}\label{dmb}
 \frac{M}{b}&=\frac{\tan\vartheta_\bullet}{\sin\vartheta}=\frac{\tan(4D\varepsilon^2)}{\sin(4D\theta\varepsilon)}\,.
\end{align}
By this way, we arrive at the aim of expressing $\hat{\alpha}$ as a function $\hat{\alpha}(\varepsilon)$ of $\varepsilon$. Up to now, we have expressed all variables in  Eq. \eqref{eq:17} as functions of $\varepsilon$. The unknown quantity $\theta$, as a function $\theta(\varepsilon)$ of $\varepsilon$, is determined by Eq. \eqref{eq:17} after inserting the expressions of ${\cal B}(\varepsilon)$, ${\cal \vartheta}(\varepsilon)$, and $\hat{\alpha}(\varepsilon)$. It turns out that the $n$-th accuracy order of $\hat{\alpha}(\varepsilon)$ in $\varepsilon$ can only determine the $n-1$-th accuracy order of $\theta(\varepsilon)$. With the correction of $\hat{\alpha}(\varepsilon)$ to fourth order on hand, we expand $\theta(\varepsilon)$ to third order as
\begin{align}\label{eq:22}
\theta(\varepsilon)=\theta_{0}+\theta_{1}\varepsilon+\theta_{2}\varepsilon^{2}+\theta_{3}\varepsilon^{3}+O(\varepsilon^4).
\end{align}
Combining Eqs. \eqref{eq:variables_function}, \eqref{dmb} and \eqref{eq:22}, and expanding the terms in Eq. \eqref{eq:17} as series in $\varepsilon$, and then we obtain the following coefficients $\theta_i$ by solving the equation \eqref{eq:17} term by term
\begin{align}\label{eq:23}
 \theta_0=&\frac{1}{2}\left(\beta\pm\sqrt{\beta^2+A_1}\right)\,,\\
 \theta_1=&\frac{A_2}{A_1+4\theta_0^2}\,, \\
 \theta_2=&\frac{1}{3\theta_0\left(A_1+4\theta_0^2\right)^3}\left\{A_1\Big[A_1^4 (1-D^2)+3 A_3 A_1-3 A_2^2\Big] \right. \notag \\
 &+4 \left(A_1^4 (D-2)(D-1)+6 A_3 A_1-6 A_2^2\right)\theta_0^2+8 \left(A_1^3 [2 \right. \notag \\
 &+D(11D-12)]+6 A_3 \Big)\theta_0^4 +64 A_1^2 D (4 D-3) \theta_0^6 \notag \\
 &+128 A_1 D^2 \theta_0^8 \Big\}\,, \\
 \theta_3=&\frac{1}{3 \theta_0^2 \left(A_1+4 \theta_0^2\right)^5}\left\{3 A_1^2 \left(2 A_2^3-3 A_1 A_3 A_2+A_1^2 A_4\right) \right.\notag \\
 &+4 A_1 \Big[A_2 \left(A_1^4 (2D-1)(D-1)-33 A_3 A_1+21 A_2^2\right) \notag \\
 &+12 A_4 A_1^2\Big]\theta_0^2+8 \Big[A_2 \left(A_1^4 [D (11D-18)+10]-78 A_3 A_1 \right. \notag \\
 &+42 A_2^2\Big)+36 A_1^2 A_4\Big]\theta_0^4+32  \Big[A_2 \left(A_1^3 [5D (5D-6)+14]\right. \notag \\
 &-30 A_3\Big)+24 A_1 A_4\Big]\theta_0^6+128  \left(A_1^2 A_2 [5D (7 D-6)+6]\right. \notag \\
 &+6 A_4\Big)\theta_0^8+512 A_1 A_2 D (19 D-12) \theta_0^{10}+4096 A_2 D^2 \theta_0^{12}\Big\}\,\label{eq:theta}.
\end{align}

The two solutions in Eq. \eqref{eq:23} imply that two images, namely the primary and secondary images, will arise for a source. Since we can not remove the lens, the angle $\beta$ (or ${\cal B}$) can not be observed directly. Without loss of generality, we assume $\beta>0$ (or ${\cal B}>0$). Denote the two solutions by
\begin{align}\label{theta0-beta}
\theta^{\pm}_0\equiv\frac12\left(\beta\pm\sqrt{\beta^2+A_1}\right)\,. 
\end{align}
$\theta^{+}_0$ and $\theta^{-}_0$ correspond to the angular positions of the primary and secondary images, respectively. We eliminate the unknown $\beta$ by multiplying $\theta^{+}_0$ and $\theta^{-}_0$, and get the more useful relation
\begin{align}
 &\theta^{+}_0\theta^{-}_0=-\frac{A_1}{4}\,.
\end{align}
The corresponding relation can be written in terms of the observable variable $\vartheta$ as
\begin{align}
 \vartheta^{+}_0\vartheta^{-}_0=-\frac{A_1}{4}\vartheta_E^2\,.
\end{align}
On the other hand, if the two angles $\theta^{\pm}_0$ (or $\vartheta^{\pm}_0$) are measured, then the angles $\beta$ and ${\cal B}$ can be calculated by
\begin{align}
 \beta&=\theta^{+}_0+\theta^{-}_0\,,\\
 {\cal B}&=\vartheta^{+}_0+\vartheta^{-}_0\,.
\end{align}
Moreover, we have the following relations
\begin{align}
 \theta^{+}_1+\theta^{-}_1&=\frac{A_2}{A_1}\,,\\
 \theta^{+}_2+\theta^{-}_2&=\frac{2\beta}{3 A_1^3}\left[A_1^4 \left(3 D^2-2\right)-6 A_3 A_1+6 A_2^2\right]\,,\\
 \theta^{+}_3+\theta^{-}_3&=\frac{1}{3 A_1^5}\left\{12 A_4 A_1^2 \left(A_1+4 \beta ^2\right)+4 A_2 \left[3 A_1 \left(A_2^2-12 A_3 \beta ^2\right)\right.\right.\notag\\
 &\hspace{0.5cm}\left.\left.+24 A_2^2 \beta ^2+A_1^5\Big(3 (D-1) D+1\Big)-6 A_3 A_1^2\right]\right\}\,.
\end{align}

The magnification $\mu$ of a lensed image is defined by the ratio between the solid angles of the image and the source as \cite{Wambsganss:1998gg,Virbhadra:1999nm}
\begin{align}\label{eq:27}
\mu(\vartheta)=\left[\frac{\sin {\cal B}(\vartheta)}{\sin(\vartheta)}\frac{\rm d {\cal B}(\vartheta)}{\rm d\vartheta} \right]^{-1}\,.
\end{align}
To calculate it, we take the derivative of the lens equation \eqref{eq:17} with respect to $\vartheta$, and get
\begin{align}\label{dbetatheta}
 \frac{\rm d {\cal B}(\vartheta)}{\rm d\vartheta}&=\left(\frac{\sec\vartheta}{\sec{\cal B}}\right)^2-D\left[\left(\frac{\sec\vartheta}{\sec{\cal B}}\right)^2+\left(\frac{\sec(\hat{\alpha}-\vartheta)}{\sec{\cal B}}\right)^2\left(\frac{{\rm d}\hat{\alpha}}{{\rm d}\vartheta}-1\right)\right]\,.
\end{align}
On the other hand, we have
\begin{align}\label{datheta}
 \frac{{\rm d}\hat{\alpha}}{{\rm d}\vartheta}&=\frac{{\rm d}\hat{\alpha}}{{\rm d}b}\frac{{\rm d}b}{{\rm d}\vartheta}=\cot\vartheta\left(b\frac{{\rm d}\hat{\alpha}}{{\rm d}b}\right)=\cot\vartheta\sum_k(-k)A_k\left(\frac{M}{b}\right)^k\,.
\end{align}
Combining Eqs. \eqref{eq:variables_function}, \eqref{dmb}, \eqref{dbetatheta} with Eq. \eqref{datheta}, we can express the magnification $\mu$ in Eq. \eqref{eq:27} as a function $\mu(\beta,\theta_i,\varepsilon)$ of $\beta$, $\theta_i$ and $\varepsilon$. We first make a series expansion in $\varepsilon$ for $\mu$, then substitute for the values of $\theta_1,\theta_2$ and $\theta_3$, and replace $\beta$ by $\frac{4(\theta_0^\pm)^2-A_1}{4\theta_0^\pm}$ for $\theta_0^\pm$ to reduce $\mu(\beta,\theta_i,\varepsilon)$ as $\mu(\theta_0,\varepsilon)$
\begin{align}\label{eq:28}
 \mu=\mu_0+\mu_1\varepsilon+\mu_2\varepsilon^2+\mu_3\varepsilon^3+O(\varepsilon^4)\,,
\end{align}
where
\begin{align}
 \mu_0=&\frac{16 \theta_0^4}{16 \theta_0^4-A_1^2}\,, \\
 \mu_1=&-\frac{16 A_2 \theta_0^3}{\left(A_1+4 \theta_0^2\right)^3}\,, \\
 \mu_2=&\frac{8 \theta_0^2}{3 \left(A_1-4 \theta_0^2\right) \left(A_1+4 \theta_0^2\right)^5}\left\{-A_1^6 D^2+8 A_1^2 \Big[A_1^3 \left(-9 D^2+6 D \right.\right. \notag \\
 &+2\Big)+6 A_3\Big]\theta_0^2-32  \left(A_1^4 [D (17 D-12)-4]-12 A_3 A_1\right. \notag \\
 &+18 A_2^2\Big)\theta_0^4-128  \left[A_1^3 \left(9 D^2-6 D-2\right)-6 A_3\right]\theta_0^6 \notag \\
 &-256 A_1^2 D^2 \theta_0^8\Big\}\,,\\
 \mu_3=&\frac{8 \theta_0}{3 \left(A_1-4 \theta_0^2\right) \left(A_1+4 \theta_0^2\right)^7}\left\{A_2 A_1^3 \Big[A_1^4 \left(2-3 D^2\right)-12 A_3 A_1 \right.\notag \\
 &+6 A_2^2\Big]+6 A_4 A_1^5+4 A_1^2 \Big[A_2 \left(A_1^4 [3 D (4-9 D)+14] \right. \notag \\
 &-84 A_3 A_1+42 A_2^2\Big)+42 A_4 A_1^2\Big]\theta_0^2-96 A_1  \Big[A_2 \left(A_1^4 (5 D^2-4)\right. \notag \\
 &+42 A_3 A_1-21 A_2^2\Big)-18 A_1^2 A_4\Big]\theta_0^4-128\Big[A_2 \left(2 A_1^4 (9 D^2\right. \notag \\
 &+3 D-8)+150 A_3 A_1-105 A_2^2\Big)-66 A_1^2 A_4\Big]\theta_0^6-256  \Big[A_2 \notag \\
 &\times\left(A_1^3 [3 D (23 D-4)-38]+120 A_3\right)-78 A_1 A_4\Big]\theta_0^8-3072 \notag \\
 &\times\left(A_1^2 A_2 (11 D^2-6)-6 A_4\right)\theta_0^{10}+24576 A_1 A_2 (D-2) D \theta_0^{12}\Big\}\,.
\end{align}
The magnification takes two values $\mu^+$ and $\mu^-$ corresponding to the positions of the primary and secondary images $\theta^+$ and $\theta^-$ respectively. The absolute value $|\mu^\pm|$ represents the corresponding image brightness. Some simplified relations can be obtained in the forms
\begin{align}
 \mu_0^++\mu_0^-&=1\,,\\
 \mu_1^++\mu_1^-&=0\,,\\
 \mu_2^++\mu_2^-&=0\,.
\end{align}
However, the simplified algebraic relations for higher-order corrections will not be held forever. For instance, sum over the two third-order correction terms yields
\begin{align}
 \mu_3^++\mu_3^-&=\frac{1}{\beta}\left[\frac{2A_2}{3}\left(3D^2-2\right)-\frac{4A_2^3}{A_1^4}+\frac{8A_2A_3}{A_1^3}-\frac{4A_4}{A_1^2}\right]\,,
\end{align}
in which the relation depending on the values of $A_i$ implies that it depends on the theories of gravity in general. 

In the case that the positions of primary and secondary images are very close to each other, the total magnification $\mu_{\rm tot}$ provides us another observable quantity, which is defined by
\begin{align}\label{eq:33}
 \mu_{\rm tot}=|\mu^+|+|\mu^-|\,.
\end{align}
Using Eq. \eqref{eq:28} and then using Eq. \eqref{theta0-beta} to express $\theta^\pm_0$ in terms of $\beta$, we have
\begin{align}\label{eq:35}
 \mu_{\rm tot}&=\frac{A_1+2 \beta ^2}{2 \beta  \sqrt{A_1+\beta ^2}}-\frac{A_2 }{2 \left(A_1+\beta ^2\right)^{3/2}}\varepsilon+\frac{1}{12 \beta  \left(A_1+\beta ^2\right)^{5/2}}\left\{9 A_2^2 \right. \notag \\
 &+2 \left(A_1+\beta ^2\right) \Big[2 A_1^2 \beta ^2 D^2+A_1^3 \left(9 D^2-6 D-2\right)-6 A_3\Big]\Big\}\varepsilon^2 \notag \\
 &+\frac{1}{12 A_1^4 \left(A_1+\beta ^2\right)^{7/2}}\left\{3 A_1^3 \Big[A_2 \left(A_1^4 (4+4 D-6 D^2)-60 A_3 A_1 \right.\right.\notag \\
 &+35 A_2^2\Big)+24 A_1^2 A_4\Big]+2 A_1^2 \Big[A_2 \left(A_1^4 (26+30 D-51 D^2)\right.\notag \\
 &-210 A_3 A_1+105 A_2^2\Big)+96 A_1^2 A_4\Big]\beta ^2 -4 A_1 \Big[A_2 \left(A_1^4 [3 D (9 D\right.\notag \\
 &-4)-14]+84 A_3 A_1-42 A_2^2\Big)-42 A_1^2 A_4\Big]\beta^4-8 \Big[A_2 \left(A_1^4 \right.\notag \\
 &\times(3 D^2-2)+12 A_3 A_1-6 A_2^2\Big)-6 A_1^2 A_4\Big]\beta^6\Big\}\varepsilon^3 +O(\varepsilon^4)\,.
\end{align}

\section{Light bending and GL in BD theory}\label{sec:III}

In this section, the general formalism derived in the previous section will be applied to the static, spherically symmetric and flat spacetime in BD theory, which allows us to directly write down the corresponding light bending angle and lensing observables.

In BD theory, the dynamical variables are the spacetime metric $g_{ab}$ and a scalar field $\phi$ determined by the following equations
\begin{align}
 R_{ab}-\frac12Rg_{ab}&=\frac{8\pi}{\phi}T_{ab}+\omega\phi^{-2}\left(\nabla_a\phi\nabla_b\phi-\frac12g_{ab}\nabla_c\phi\nabla^c\phi\right)\notag\\
 &\quad+\phi^{-1}\left(\nabla_a\nabla_b\phi-g_{ab}\nabla_c\nabla^c\phi\right)\,,\\
\nabla_c\nabla^c\phi&=\frac{8\pi}{3+2\omega}T\,,
\end{align}
where $\nabla_a$ is the covariant derivative operator compatible with $g_{ab}$, $R_{ab}$ is the Ricci tensor field, $R$ is the scalar curvature, $T_{ab}$ is the energy-momentum tensor of matter, $T$ is the trace of $T_{ab}$, and $\omega$ is the so-called BD parameter. On the contrary to GR, the static, spherically symmetric and asymptotically flat solution to the vacuum BD field equations is not unique. There exist four different vacuum solutions \cite{Brans:1961sx,Brans:1962zz}. However only two of them are really independent \cite{Bhadra:2005mc}. Moreover, the solution given in \cite{Brans:1961sx} is the only one satisfying some appropriate physical conditions \cite{Bhadra:2005mc}, which corresponds to the line element \eqref{eq:1} with the following coefficients \cite{Brans:1961sx}
\begin{align}
A(\rho)=&\left(\frac{1-E/\rho}{1+E/\rho}\right)^{2/\sigma}\,,\label{eq:BD-A} \\
B(\rho)=&(1+E/\rho)^{4}\left(\frac{1-E/\rho}{1+E/\rho}\right)^{2[(\sigma-C-1)/\sigma]}\,,\label{eq:BD-B}
\end{align}
where
\begin{align}\label{eq:38}
E&=\frac{M}{2}\sqrt{\frac{3+2\omega}{4+2\omega}}\,, \quad C=-\frac{1}{2+\omega}\,, \quad\sigma=\sqrt{\frac{3+2\omega}{4+2\omega}}\,.
\end{align}
It is easy to see that the solution makes sense only for the BD parameter $\omega<-2$ or $\omega>-\frac32$. The line element recovers the Schwarzschild line element in the isotropic coordinates when $\omega\rightarrow\infty$, which reflects the fact that BD theory can be regarded as a natural generalization of GR.

\subsection{Bending angle in BD theory}

Making Taylor series expansion in powers of $M/\rho$ for $A(\rho)$ and $B(\rho)$ given in Eqs. \eqref{eq:BD-A} and \eqref{eq:BD-B}, one obtains the corresponding PPN parameters as follows:
\begin{align}\label{eq:39}
\alpha&=1\,, \qquad \lambda=1\,,\qquad \xi=\frac{18 \omega +35}{18 (\omega +2)}\,,\qquad \kappa=\frac{6 \omega +11}{6 (\omega +2)}\,,\notag\\
\gamma&=\frac{\omega +1}{\omega+2}\,,\delta=\frac{6 \omega ^2+9\omega+2}{6(\omega+2)^2}\,,\eta=\frac{6 \omega^3+3\omega^2-17\omega-14}{6(\omega+2)^3}\,,\notag\\
   \nu&=\frac{12\omega^4-108\omega^3-421\omega^2-452\omega-148}{12(\omega+2)^4}\,.
\end{align}
Putting these PPN parameters into Eq. \eqref{eq:16} yields the light bending angle $\hat{\alpha}^{\st BD}(b)$ for BD theory
\begin{align}\label{eq:BD-bending}
\hat{\alpha}^{\st BD}(b)=&A^{\st BD}_{1}\frac{M}{b}+A^{\st BD}_{2}\left(\frac{M}{b}\right)^2+A^{\st BD}_{3}\left(\frac{M}{b}\right)^3\notag\\
         &+A^{\st BD}_{4}\left(\frac{M}{b}\right)^4+O\left[\left(\frac{M}{b}\right)^5\right]\,,
\end{align}
with
\begin{align}\label{eq:40}
A^{\st BD}_{1}&=\frac{4\omega+6}{\omega +2}\,,\qquad A^{\st BD}_{2}=\frac{\pi\left(30\omega^2+89\omega+66\right)}{8(\omega+2)^2}\,,\notag\\
A^{\st BD}_{3}&=\frac{2(2\omega +3)^2(16\omega+23)}{3(\omega+2)^3}\,,\notag\\
A^{\st BD}_{4}&=\frac{\pi(2\omega+3)^2\left(3465\omega^2+10084\omega+7332\right)}{256(\omega+2)^4}\,.
\end{align}

\subsection{Positions and Magnifications of the lensing images in BD theory}

With the bending angle $\hat{\alpha}^{\st BD}(b)$ in Eq. \eqref{eq:BD-bending}, we can obtain the positions $\theta(\varepsilon)$ of the lensing image, by plugging the coefficients in Eq. \eqref{eq:40} into Eq. \eqref{eq:22}, with the expanding coefficients
\begin{align}\label{eq:41}
\theta^{\st BD}_{0} =&\frac{1}{2}\left(\beta\pm\sqrt{\beta^{2}+4\frac{3+2\omega}{4+2\omega}}\right)\,, \\
\theta^{\st BD}_{1} =&\frac{\pi\left(30\omega^2+89\omega+66\right)}{16 (\omega +2)\left[2\theta_0^2(\omega+2)+2\omega+3\right]}\,,\\
\theta^{\st BD}_{2} =&\frac{2\omega+3}{768\theta_0(\omega+2)^2\left[2\theta_0^2(\omega+2)+2\omega+3\right]^3}\left\{-(2\omega +3)^2\right. \notag\\
           &\times\left[1024D^2(2\omega+3)^2-256(24\omega+35)(2\omega+3)+3\pi^2\right. \notag\\
           &\times(15\omega+22)^2\Big]+4(\omega+2)(2\omega+3)\left[512D^2(2\omega+3)^2\right. \notag\\
           &-1536D(2\omega+3)^2+256(24\omega+35)(2\omega+3)-3\pi^2 \notag\\
           &\times(15\omega+22)^2\Big]\theta_0^2+1024(\omega+2)^2(2\omega+3)\Big[2D(11D \notag\\
           &-12)(2 \omega+3)+24\omega+35\Big]\theta_0^4+8192D(4D-3)\notag\\
           &\times(\omega+2)^3(2\omega+3)\theta_0^6+8192 D^2(\omega+2)^4\theta_0^8 \Big\}\,.
\end{align}
Here the expression of the third-correction $\theta^{\st BD}_3$ is not shown due to its complicated form.

\begin{figure}
  \includegraphics[width=\columnwidth]{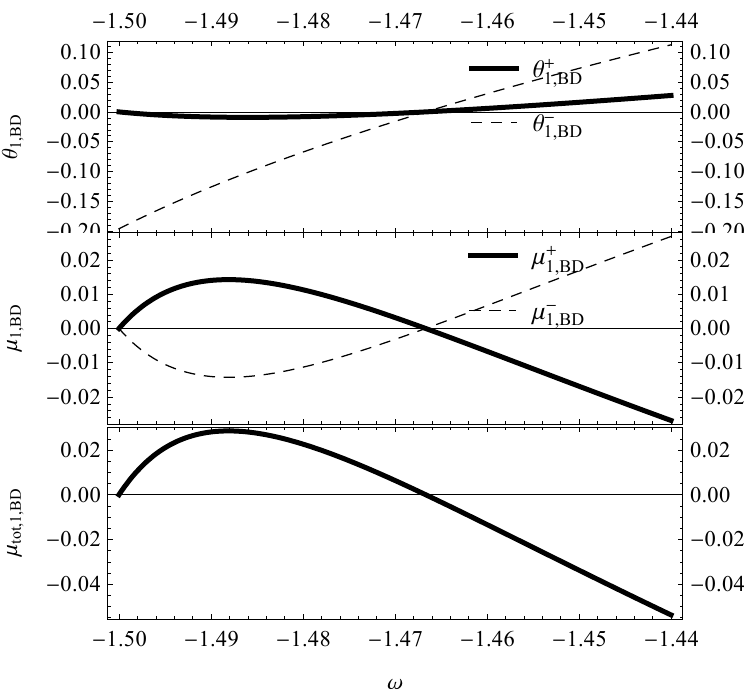}
  \caption{The first-corrections to the positions, the magnifications and the total magnification of the images as functions of $\omega$ in BD theory for $\beta=0.5$ and $D=0.5$.}
  \label{first-corrections}
\end{figure}

\begin{figure*}[t]
  \includegraphics[height=105mm]{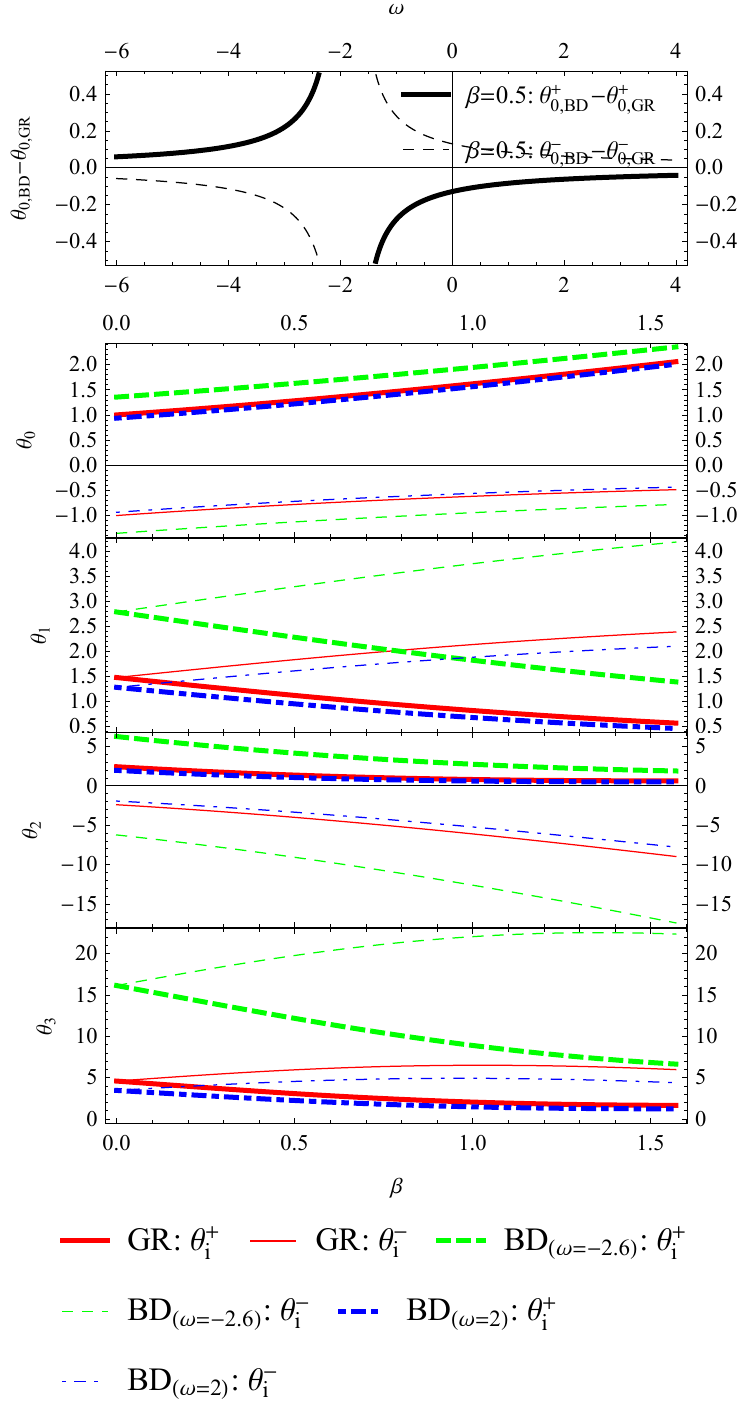}
  \hspace{8pt}
  \includegraphics[height=105mm]{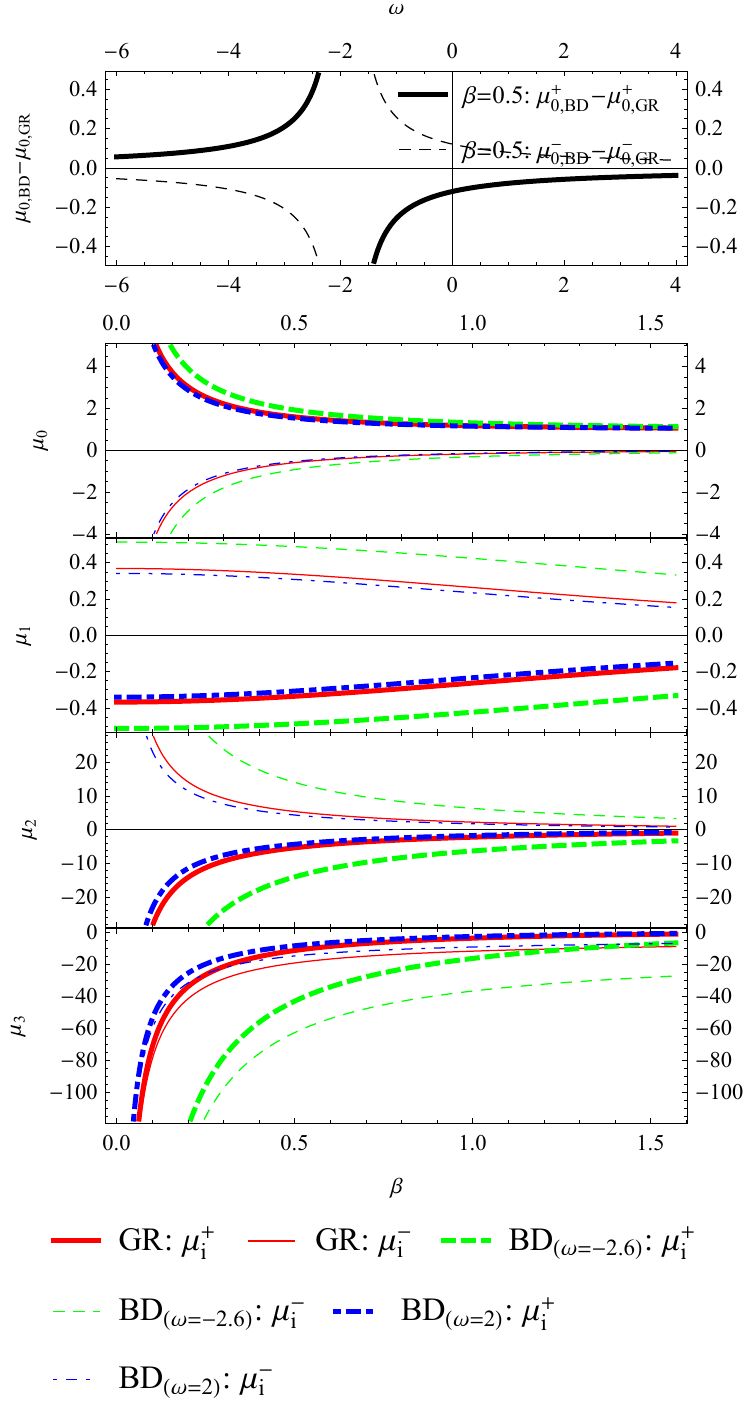}
  \hspace{8pt}
  \includegraphics[height=105mm]{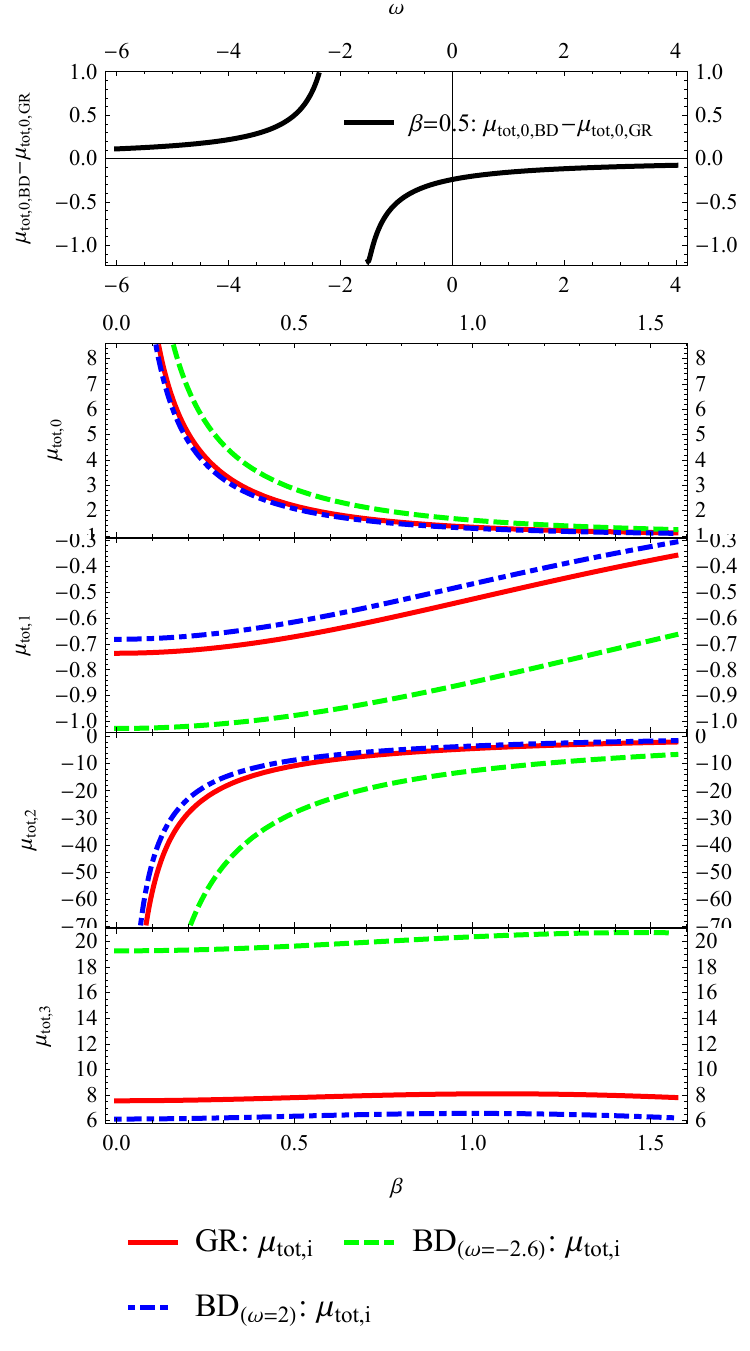}
  \caption{The differences between BD theroy and GR in corrections to the positions, the magnifications and the total magnification of the primary and the secondary images for $D=0.5$.}
  \label{fig:comparision}
\end{figure*}

Similarly, we obtain the magnifications of the lensing image $\mu(\varepsilon)$ in Eq. \eqref{eq:28} with the following coefficients
\begin{align}\label{eq:44}
 \mu^{\st BD}_0=&\frac{4 \theta_0^4(\omega +2)^2}{4 \theta_0^4(\omega+2)^2-(2\omega+3)^2}\,,\\
 \mu^{\st BD}_1=&-\frac{\pi\theta_0^3(\omega +2)(2\omega +3)(15\omega+22)}{4\left[2\theta_0^2(\omega+2)+2\omega +3\right]^3}\,,\\
 \mu^{\st BD}_2=&\frac{\theta_0^2(2\omega +3)^2}{24\left[2\theta_0^2(\omega +2)-2\omega-3\right]\left[2\theta_0^2(\omega+2)+2\omega +3\right]^5} \notag\\
 &\times\left\{64D^2(2\omega+3)^4+128(\omega+2)(2\omega+3)^2\Big[6 D (3 D-2)\right.\notag\\
 &\times(2\omega+3)-24\omega-35\Big]\theta_0^2+(\omega+2)^2\Big[8704 D^2(2\omega+3)^2\notag\\
 &-6144D(2\omega+3)^2-512(24\omega+35)(2\omega+3)+9\pi^2 \notag\\
 &\times(15\omega+22)^2\Big]\theta_0^4+512(\omega +2)^3\Big[6D(3D-2)(2\omega+3) \notag\\
 &-24\omega-35\Big]\theta_0^6+1024 D^2(\omega+2)^4\theta_0^8\Big\}\,.
\end{align}
Here the expression of $\mu^{\st BD}_3$ is also omitted because of its complicated form.

The total magnification for the BD theory reads
\begin{align}\label{eq:47}
\mu^{\st BD}_{\rm tot}=&\frac{\beta^2(\omega +2)+2\omega +3}{\beta(\omega +2)n}-\frac{\pi(2\omega+3)(15\omega +22)n}{16 \left[\beta ^2 (\omega +2)+4\omega +6\right]^2}\varepsilon \notag \\
&+\frac{(2\omega+3)^2}{768\beta(\omega+2)^2n\left[\beta ^2 (\omega+2)+4\omega+6\right]^2}\Big\{-512(24\omega+35)\notag \\
&\times\left[\beta^2(\omega+2)+4\omega+6\right]+1024 D^2\left[\beta^2(\omega+2)+4\omega +6\right]\notag \\
&\times\left[\beta^2(\omega +2)+9(2\omega +3)\right]-6144 D(2\omega+3)\left[\beta^2(\omega+2)\right. \notag \\
&+4\omega+6\Big]+9\pi^2(15\omega+22)^2\Big\}\varepsilon^2+O(\varepsilon^3)\,,
\end{align}
where $n=\sqrt{\beta^2-\frac{2}{\omega +2}+4}$\,.

\section{Summary and discussion}\label{sec:IV}

In this paper, we studies light bending and GL in BD theory. Firstly, we presented the integral formalism of bending angle $\hat{\alpha }(\rho_0)$ in Eq. \eqref{eq:3} for a light ray moving in a static, spherically symmetric and asymptotically flat spacetime in the isotropic coordinate system. We then derived, based on PPN metric in the isotropic coordinates given by Eqs. \eqref{eq:6} and \eqref{eq:7}, higher-order corrections to bending angle $\hat{\alpha}(b)$ in Eq. \eqref{eq:15} in terms of invariant impact parameter $b$ by carrying out the integration \eqref{eq:3} (or \eqref{eq:5}) term by term. Based on the expression of light bending angle, we solved the lens equation \eqref{eq:17} to get higher-order corrections to positions of the lensed images $\theta(\varepsilon)$ in \eqref{eq:22} with coefficients $\theta_i$ in Eqs. \eqref{eq:23}-\eqref{eq:theta}. The magnifications and total magnification of the lensed images were also evaluated, and given in Eqs. \eqref{eq:28} and \eqref{eq:35}. All general expressions of bending angle and lensing observables are characterized by the PPN parameters. Finally, we directly wrote down the light bending angle and the lensing observables by putting the corresponding PPN parameters of the static, spherically symmetric and asymptotically metric in BD theory into the general formalism.

A similar work has been studied for a spherically symmetric metrics expanded as a PPN series in the Schwarzschild coordinates \cite{Keeton:2005jd,Keeton:2006sa}. Since there is an analytic relation between the Schwarzschild and the isotropic coordinates for a spherically symmetric metric expanded as PPN series only to third order, the bending angle in terms of the impact parameter $b$ to third-order terms obtained in this paper coincides with the one in \cite{Keeton:2005jd,Keeton:2006sa}, although different coordinates are adopted to do calculations. The general formalism presented in this paper is suitable for calculating higher-order correction to the bending angle for a metric written in the isotropic coordinates.  Moreover, our results show that first-order correction to total magnification $\mu_{\rm tot}$ in Eq. \eqref{eq:35} does not vanish in general, rather than a trival (vanishing) contribution concluded in \cite{Keeton:2005jd,Keeton:2006sa}. The reason for the two different results obtained in this paper and in \cite{Keeton:2005jd,Keeton:2006sa} comes from different assignments for the positions of source and images. In \cite{Keeton:2005jd,Keeton:2006sa}, the angles of image positions are assumed to be positive, which leads to the position of the source to take on a positive or negative value depending on the image's location. In this paper, we fix the source position $\beta>0$ without loss of generality, which is more suitable for astronomical observation.

Finally, let us investigate the effects of the undeterminated parameter $\omega$ in BD theory in numerical analysis. The first-corrections to the positions, the magnifications and the total magnification of the images as functions of $\omega$ for $\beta=0.5$ and $D=0.5$ were plotted in Fig. \ref{first-corrections}. $-3/2<\omega<-22/15$ corresponding to $A_2<0$ in BD theory, the positive-parity image will be close to the lens, while the negative-parity image will shift away from the lens due to the first-order corrections $\theta^\pm_{1,{\rm BD}}<0$. The positive-parity, negative-parity images and the total image get brighter due to $\mu^+_{1,{\rm BD}}>0, \mu^-_{1,{\rm BD}}<0$ and $\mu_{{\rm tot},1,{\rm BD}}>0$. When $\omega>-22/15$ or $\omega<-2$ corresponding to $A_2>0$, the opposition occurs. Fig. \ref{fig:comparision} shows the difference between BD theory and GR in corrections to the positions, the magnifications and the total magnification of the primary and the secondary images. The upper three panels show that the leading-order corrections to lensing observables in BD theory close to those in GR when the parameter $\omega$ tends to $+\infty$ from $-\frac32$, while opposition occurs when $\omega$ tends to $-2$ from $-\infty$. The variation of the terms in a series expansion of lensing observables with the angular position of source $\beta$ in BD theory and GR were plotted in the other panels.

\begin{acknowledgments}
This work is supported in part by the NSFC Grants No. 11765006 and No. 11875006.
\end{acknowledgments}


%


\begin{thebibliography}{31}%
\makeatletter
\providecommand \@ifxundefined [1]{%
 \@ifx{#1\undefined}
}%
\providecommand \@ifnum [1]{%
 \ifnum #1\expandafter \@firstoftwo
 \else \expandafter \@secondoftwo
 \fi
}%
\providecommand \@ifx [1]{%
 \ifx #1\expandafter \@firstoftwo
 \else \expandafter \@secondoftwo
 \fi
}%
\providecommand \natexlab [1]{#1}%
\providecommand \enquote  [1]{#1}%
\providecommand \bibnamefont  [1]{#1}%
\providecommand \bibfnamefont [1]{#1}%
\providecommand \citenamefont [1]{#1}%
\providecommand \href@noop [0]{\@secondoftwo}%
\providecommand \href [0]{\begingroup \@sanitize@url \@href}%
\providecommand \@href[1]{\@@startlink{#1}\@@href}%
\providecommand \@@href[1]{\endgroup#1\@@endlink}%
\providecommand \@sanitize@url [0]{\catcode `\\12\catcode `\$12\catcode
  `\&12\catcode `\#12\catcode `\^12\catcode `\_12\catcode `\%12\relax}%
\providecommand \@@startlink[1]{}%
\providecommand \@@endlink[0]{}%
\providecommand \url  [0]{\begingroup\@sanitize@url \@url }%
\providecommand \@url [1]{\endgroup\@href {#1}{\urlprefix }}%
\providecommand \urlprefix  [0]{URL }%
\providecommand \Eprint [0]{\href }%
\providecommand \doibase [0]{http://dx.doi.org/}%
\providecommand \selectlanguage [0]{\@gobble}%
\providecommand \bibinfo  [0]{\@secondoftwo}%
\providecommand \bibfield  [0]{\@secondoftwo}%
\providecommand \translation [1]{[#1]}%
\providecommand \BibitemOpen [0]{}%
\providecommand \bibitemStop [0]{}%
\providecommand \bibitemNoStop [0]{.\EOS\space}%
\providecommand \EOS [0]{\spacefactor3000\relax}%
\providecommand \BibitemShut  [1]{\csname bibitem#1\endcsname}%
\let\auto@bib@innerbib\@empty
\bibitem [{\citenamefont {Brans}\ and\ \citenamefont
  {Dicke}(1961)}]{Brans:1961sx}%
  \BibitemOpen
  \bibfield  {author} {\bibinfo {author} {\bibfnamefont {C.}~\bibnamefont
  {Brans}}\ and\ \bibinfo {author} {\bibfnamefont {R.~H.}\ \bibnamefont
  {Dicke}},\ }\bibfield  {title} {\enquote {\bibinfo {title} {{Mach's principle
  and a relativistic theory of gravitation}},}\ }\href {\doibase
  10.1103/PhysRev.124.925} {\bibfield  {journal} {\bibinfo  {journal} {Phys.
  Rev.}\ }\textbf {\bibinfo {volume} {124}},\ \bibinfo {pages} {925--935}
  (\bibinfo {year} {1961})}\BibitemShut {NoStop}%
\bibitem [{\citenamefont {Will}(2014)}]{Will:2014kxa}%
  \BibitemOpen
  \bibfield  {author} {\bibinfo {author} {\bibfnamefont {C.~M.}\ \bibnamefont
  {Will}},\ }\bibfield  {title} {\enquote {\bibinfo {title} {{The confrontation
  between general relativity and experiment}},}\ }\href {\doibase
  10.12942/lrr-2014-4} {\bibfield  {journal} {\bibinfo  {journal} {Living Rev.
  Rel.}\ }\textbf {\bibinfo {volume} {17}},\ \bibinfo {pages} {4} (\bibinfo
  {year} {2014})},\ \Eprint {http://arxiv.org/abs/1403.7377} {arXiv:1403.7377
  [gr-qc]}\BibitemShut {NoStop}%
\bibitem [{\citenamefont {Bertolami}\ and\ \citenamefont
  {Martins}(2000)}]{Bertolami:1999dp}%
  \BibitemOpen
  \bibfield  {author} {\bibinfo {author} {\bibfnamefont {O.}~\bibnamefont
  {Bertolami}}\ and\ \bibinfo {author} {\bibfnamefont {P.~J.}\ \bibnamefont
  {Martins}},\ }\bibfield  {title} {\enquote {\bibinfo {title} {{Nonminimal
  coupling and quintessence}},}\ }\href {\doibase 10.1103/PhysRevD.61.064007}
  {\bibfield  {journal} {\bibinfo  {journal} {Phys. Rev. D}\ }\textbf {\bibinfo
  {volume} {61}},\ \bibinfo {pages} {064007} (\bibinfo {year} {2000})},\
  \Eprint {http://arxiv.org/abs/gr-qc/9910056} {arXiv:gr-qc/9910056
  [gr-qc]}\BibitemShut {NoStop}%
\bibitem [{\citenamefont {Qiang}\ \emph {et~al.}(2005)\citenamefont {Qiang},
  \citenamefont {Ma}, \citenamefont {Han},\ and\ \citenamefont
  {Yu}}]{Qiang:2004gg}%
  \BibitemOpen
  \bibfield  {author} {\bibinfo {author} {\bibfnamefont {L.}~\bibnamefont
  {Qiang}}, \bibinfo {author} {\bibfnamefont {Y.}~\bibnamefont {Ma}}, \bibinfo
  {author} {\bibfnamefont {M.}~\bibnamefont {Han}}, and\ \bibinfo {author}
  {\bibfnamefont {D.}~\bibnamefont {Yu}},\ }\bibfield  {title} {\enquote
  {\bibinfo {title} {{Five-dimensional Brans-Dicke theory and cosmic
  acceleration}},}\ }\href {\doibase 10.1103/PhysRevD.71.061501} {\bibfield
  {journal} {\bibinfo  {journal} {Phys. Rev. D}\ }\textbf {\bibinfo {volume}
  {71}},\ \bibinfo {pages} {061501} (\bibinfo {year} {2005})},\ \Eprint
  {http://arxiv.org/abs/gr-qc/0411066} {arXiv:gr-qc/0411066
  [gr-qc]}\BibitemShut {NoStop}%
\bibitem [{\citenamefont {Bisabr}(2012)}]{Bisabr:2011re}%
  \BibitemOpen
  \bibfield  {author} {\bibinfo {author} {\bibfnamefont {Y.}~\bibnamefont
  {Bisabr}},\ }\bibfield  {title} {\enquote {\bibinfo {title} {{Cosmic
  acceleration in Brans-Dicke cosmology}},}\ }\href {\doibase
  10.1007/s10714-011-1281-8} {\bibfield  {journal} {\bibinfo  {journal} {Gen.
  Rel. Grav.}\ }\textbf {\bibinfo {volume} {44}},\ \bibinfo {pages} {427--435}
  (\bibinfo {year} {2012})},\ \Eprint {http://arxiv.org/abs/1110.3421}
  {arXiv:1110.3421 [gr-qc]}\BibitemShut {NoStop}%
\bibitem [{\citenamefont {Einstein}(1916)}]{Einstein:1916vd}%
  \BibitemOpen
  \bibfield  {author} {\bibinfo {author} {\bibfnamefont {A.}~\bibnamefont
  {Einstein}},\ }\bibfield  {title} {\enquote {\bibinfo {title} {{Die Grundlage
  der allgemeinen Relativit{\"a}tstheorie}},}\ }\href {\doibase
  10.1002/andp.19163540702} {\bibfield  {journal} {\bibinfo  {journal} {Annalen
  der Physik}\ }\textbf {\bibinfo {volume} {49}},\ \bibinfo {pages} {769--822}
  (\bibinfo {year} {1916})}\BibitemShut {NoStop}%
\bibitem [{\citenamefont {Dyson}\ \emph {et~al.}(1920)\citenamefont {Dyson},
  \citenamefont {Eddington},\ and\ \citenamefont {Davidson}}]{Dyson:1920cwa}%
  \BibitemOpen
  \bibfield  {author} {\bibinfo {author} {\bibfnamefont {F.~W.}\ \bibnamefont
  {Dyson}}, \bibinfo {author} {\bibfnamefont {A.~S.}\ \bibnamefont
  {Eddington}}, and\ \bibinfo {author} {\bibfnamefont {C.}~\bibnamefont
  {Davidson}},\ }\bibfield  {title} {\enquote {\bibinfo {title} {{A
  determination of the deflection of light by the Sun's gravitational field,
  from observations made at the total eclipse of May 29, 1919}},}\ }\href
  {\doibase 10.1098/rsta.1920.0009} {\bibfield  {journal} {\bibinfo  {journal}
  {Phil. Trans. Roy. Soc. Lond. A}\ }\textbf {\bibinfo {volume} {220}},\
  \bibinfo {pages} {291--333} (\bibinfo {year} {1920})}\BibitemShut {NoStop}%
\bibitem [{\citenamefont {Lebach}\ \emph {et~al.}(1995)\citenamefont {Lebach},
  \citenamefont {Corey}, \citenamefont {Shapiro}, \citenamefont {Ratner},
  \citenamefont {Webber}, \citenamefont {Rogers}, \citenamefont {Davis},\ and\
  \citenamefont {Herring}}]{Lebach:1995zz}%
  \BibitemOpen
  \bibfield  {author} {\bibinfo {author} {\bibfnamefont {D.~E.}\ \bibnamefont
  {Lebach}}, \bibinfo {author} {\bibfnamefont {B.~E.}\ \bibnamefont {Corey}},
  \bibinfo {author} {\bibfnamefont {I.~I.}\ \bibnamefont {Shapiro}}, \bibinfo
  {author} {\bibfnamefont {M.~I.}\ \bibnamefont {Ratner}}, \bibinfo {author}
  {\bibfnamefont {J.~C.}\ \bibnamefont {Webber}}, \bibinfo {author}
  {\bibfnamefont {A.~E.~E.}\ \bibnamefont {Rogers}}, \bibinfo {author}
  {\bibfnamefont {J.~L.}\ \bibnamefont {Davis}}, and\ \bibinfo {author}
  {\bibfnamefont {T.~A.}\ \bibnamefont {Herring}},\ }\bibfield  {title}
  {\enquote {\bibinfo {title} {{Measurement of the Solar gravitational
  deflection of radio waves using very-long-baseline interferometry}},}\ }\href
  {\doibase 10.1103/PhysRevLett.75.1439} {\bibfield  {journal} {\bibinfo
  {journal} {Phys. Rev. Lett.}\ }\textbf {\bibinfo {volume} {75}},\ \bibinfo
  {pages} {1439--1442} (\bibinfo {year} {1995})}\BibitemShut {NoStop}%
\bibitem [{\citenamefont {Einstein}(1936)}]{Einstein:1956zz}%
  \BibitemOpen
  \bibfield  {author} {\bibinfo {author} {\bibfnamefont {A.}~\bibnamefont
  {Einstein}},\ }\bibfield  {title} {\enquote {\bibinfo {title} {{Lens-like
  action of a star by the deviation of light in the gravitational field}},}\
  }\href {\doibase 10.1126/science.84.2188.506} {\bibfield  {journal} {\bibinfo
   {journal} {Science}\ }\textbf {\bibinfo {volume} {84}},\ \bibinfo {pages}
  {506--507} (\bibinfo {year} {1936})}\BibitemShut {NoStop}%
\bibitem [{\citenamefont {Zwicky}(1937)}]{Zwicky:1937zzb}%
  \BibitemOpen
  \bibfield  {author} {\bibinfo {author} {\bibfnamefont {F.}~\bibnamefont
  {Zwicky}},\ }\bibfield  {title} {\enquote {\bibinfo {title} {{Nebulae as
  gravitational lenses}},}\ }\href {\doibase 10.1103/PhysRev.51.290} {\bibfield
   {journal} {\bibinfo  {journal} {Phys. Rev.}\ }\textbf {\bibinfo {volume}
  {51}},\ \bibinfo {pages} {290} (\bibinfo {year} {1937})}\BibitemShut
  {NoStop}%
\bibitem [{\citenamefont {Schneider}\ \emph {et~al.}(1992)\citenamefont
  {Schneider}, \citenamefont {Ehlers},\ and\ \citenamefont
  {Falco}}]{Schneider:1992bk}%
  \BibitemOpen
  \bibfield  {author} {\bibinfo {author} {\bibfnamefont {P.}~\bibnamefont
  {Schneider}}, \bibinfo {author} {\bibfnamefont {J.}~\bibnamefont {Ehlers}},
  and\ \bibinfo {author} {\bibfnamefont {E.~E.}\ \bibnamefont {Falco}},\ }\href
  {\doibase 10.1007/978-1-4612-2756-4} {\emph {\bibinfo {title} {{Gravitational
  Lenses}}}}\ (\bibinfo  {publisher} {Springer, New York},\ \bibinfo {year}
  {1992})\BibitemShut {NoStop}%
\bibitem [{\citenamefont {Wambsganss}(1998)}]{Wambsganss:1998gg}%
  \BibitemOpen
  \bibfield  {author} {\bibinfo {author} {\bibfnamefont {J.}~\bibnamefont
  {Wambsganss}},\ }\bibfield  {title} {\enquote {\bibinfo {title}
  {{Gravitational lensing in astronomy}},}\ }\href {\doibase
  10.12942/lrr-1998-12} {\bibfield  {journal} {\bibinfo  {journal} {Living Rev.
  Rel.}\ }\textbf {\bibinfo {volume} {1}},\ \bibinfo {pages} {12} (\bibinfo
  {year} {1998})},\ \Eprint {http://arxiv.org/abs/astro-ph/9812021}
  {arXiv:astro-ph/9812021 [astro-ph]}\BibitemShut {NoStop}%
\bibitem [{\citenamefont {Perlick}(2004)}]{Perlick:2004tq}%
  \BibitemOpen
  \bibfield  {author} {\bibinfo {author} {\bibfnamefont {V.}~\bibnamefont
  {Perlick}},\ }\bibfield  {title} {\enquote {\bibinfo {title} {{Gravitational
  lensing from a spacetime perspective}},}\ }\href {\doibase
  10.12942/lrr-2004-9} {\bibfield  {journal} {\bibinfo  {journal} {Living Rev.
  Rel.}\ }\textbf {\bibinfo {volume} {7}},\ \bibinfo {pages} {9} (\bibinfo
  {year} {2004})}\BibitemShut {NoStop}%
\bibitem [{\citenamefont {Bozza}(2010)}]{Bozza:2009yw}%
  \BibitemOpen
  \bibfield  {author} {\bibinfo {author} {\bibfnamefont {V.}~\bibnamefont
  {Bozza}},\ }\bibfield  {title} {\enquote {\bibinfo {title} {{Gravitational
  lensing by black holes}},}\ }\href {\doibase 10.1007/s10714-010-0988-2}
  {\bibfield  {journal} {\bibinfo  {journal} {Gen. Rel. Grav.}\ }\textbf
  {\bibinfo {volume} {42}},\ \bibinfo {pages} {2269--2300} (\bibinfo {year}
  {2010})},\ \Eprint {http://arxiv.org/abs/0911.2187} {arXiv:0911.2187
  [gr-qc]}\BibitemShut {NoStop}%
\bibitem [{\citenamefont {Walsh}\ \emph {et~al.}(1979)\citenamefont {Walsh},
  \citenamefont {Carswell},\ and\ \citenamefont {Weymann}}]{Walsh:1979nx}%
  \BibitemOpen
  \bibfield  {author} {\bibinfo {author} {\bibfnamefont {D.}~\bibnamefont
  {Walsh}}, \bibinfo {author} {\bibfnamefont {R.~F.}\ \bibnamefont {Carswell}},
  and\ \bibinfo {author} {\bibfnamefont {R.~J.}\ \bibnamefont {Weymann}},\
  }\bibfield  {title} {\enquote {\bibinfo {title} {{0957+561 A, B: twin
  quasistellar objects or gravitational lens?}}}\ }\href {\doibase
  10.1038/279381a0} {\bibfield  {journal} {\bibinfo  {journal} {Nature}\
  }\textbf {\bibinfo {volume} {279}},\ \bibinfo {pages} {381--384} (\bibinfo
  {year} {1979})}\BibitemShut {NoStop}%
\bibitem [{\citenamefont {Dey}(2012)}]{Dey:2012ph}%
  \BibitemOpen
  \bibfield  {author} {\bibinfo {author} {\bibfnamefont {T.~K.}\ \bibnamefont
  {Dey}},\ }\bibfield  {title} {\enquote {\bibinfo {title} {{Strong
  gravitational lensing by Schwarzschild black hole}},}\ }\href@noop {}
  {}\Eprint {http://arxiv.org/abs/1208.3306} {arXiv:1208.3306
  [gr-qc]}\BibitemShut {NoStop}%
\bibitem [{\citenamefont {Bozza}\ \emph {et~al.}(2001)\citenamefont {Bozza},
  \citenamefont {Capozziello}, \citenamefont {Iovane},\ and\ \citenamefont
  {Scarpetta}}]{Bozza:2001xd}%
  \BibitemOpen
  \bibfield  {author} {\bibinfo {author} {\bibfnamefont {V.}~\bibnamefont
  {Bozza}}, \bibinfo {author} {\bibfnamefont {S.}~\bibnamefont {Capozziello}},
  \bibinfo {author} {\bibfnamefont {G.}~\bibnamefont {Iovane}}, and\ \bibinfo
  {author} {\bibfnamefont {G.}~\bibnamefont {Scarpetta}},\ }\bibfield  {title}
  {\enquote {\bibinfo {title} {{Strong field limit of black hole gravitational
  lensing}},}\ }\href {\doibase 10.1023/A:1012292927358} {\bibfield  {journal}
  {\bibinfo  {journal} {Gen. Rel. Grav.}\ }\textbf {\bibinfo {volume} {33}},\
  \bibinfo {pages} {1535--1548} (\bibinfo {year} {2001})},\ \Eprint
  {http://arxiv.org/abs/gr-qc/0102068} {arXiv:gr-qc/0102068
  [gr-qc]}\BibitemShut {NoStop}%
\bibitem [{\citenamefont {Eiroa}\ \emph {et~al.}(2002)\citenamefont {Eiroa},
  \citenamefont {Romero},\ and\ \citenamefont {Torres}}]{Eiroa:2002mk}%
  \BibitemOpen
  \bibfield  {author} {\bibinfo {author} {\bibfnamefont {E.~F.}\ \bibnamefont
  {Eiroa}}, \bibinfo {author} {\bibfnamefont {G.~E.}\ \bibnamefont {Romero}},
  and\ \bibinfo {author} {\bibfnamefont {D.~F.}\ \bibnamefont {Torres}},\
  }\bibfield  {title} {\enquote {\bibinfo {title} {{Reissner-Nordstrom black
  hole lensing}},}\ }\href {\doibase 10.1103/PhysRevD.66.024010} {\bibfield
  {journal} {\bibinfo  {journal} {Phys. Rev. D}\ }\textbf {\bibinfo {volume}
  {66}},\ \bibinfo {pages} {024010} (\bibinfo {year} {2002})},\ \Eprint
  {http://arxiv.org/abs/gr-qc/0203049} {arXiv:gr-qc/0203049
  [gr-qc]}\BibitemShut {NoStop}%
\bibitem [{\citenamefont {Bozza}\ \emph {et~al.}(2005)\citenamefont {Bozza},
  \citenamefont {De~Luca}, \citenamefont {Scarpetta},\ and\ \citenamefont
  {Sereno}}]{Bozza:2005tg}%
  \BibitemOpen
  \bibfield  {author} {\bibinfo {author} {\bibfnamefont {V.}~\bibnamefont
  {Bozza}}, \bibinfo {author} {\bibfnamefont {F.}~\bibnamefont {De~Luca}},
  \bibinfo {author} {\bibfnamefont {G.}~\bibnamefont {Scarpetta}}, and\
  \bibinfo {author} {\bibfnamefont {M.}~\bibnamefont {Sereno}},\ }\bibfield
  {title} {\enquote {\bibinfo {title} {{Analytic Kerr black hole lensing for
  equatorial observers in the strong deflection limit}},}\ }\href {\doibase
  10.1103/PhysRevD.72.083003} {\bibfield  {journal} {\bibinfo  {journal} {Phys.
  Rev. D}\ }\textbf {\bibinfo {volume} {72}},\ \bibinfo {pages} {083003}
  (\bibinfo {year} {2005})},\ \Eprint {http://arxiv.org/abs/gr-qc/0507137}
  {arXiv:gr-qc/0507137 [gr-qc]}\BibitemShut {NoStop}%
\bibitem [{\citenamefont {Bozza}\ \emph {et~al.}(2006)\citenamefont {Bozza},
  \citenamefont {De~Luca},\ and\ \citenamefont {Scarpetta}}]{Bozza:2006nm}%
  \BibitemOpen
  \bibfield  {author} {\bibinfo {author} {\bibfnamefont {V.}~\bibnamefont
  {Bozza}}, \bibinfo {author} {\bibfnamefont {F.}~\bibnamefont {De~Luca}}, and\
  \bibinfo {author} {\bibfnamefont {G.}~\bibnamefont {Scarpetta}},\ }\bibfield
  {title} {\enquote {\bibinfo {title} {{Kerr black hole lensing for generic
  observers in the strong deflection limit}},}\ }\href {\doibase
  10.1103/PhysRevD.74.063001} {\bibfield  {journal} {\bibinfo  {journal} {Phys.
  Rev. D}\ }\textbf {\bibinfo {volume} {74}},\ \bibinfo {pages} {063001}
  (\bibinfo {year} {2006})},\ \Eprint {http://arxiv.org/abs/gr-qc/0604093}
  {arXiv:gr-qc/0604093 [gr-qc]}\BibitemShut {NoStop}%
\bibitem [{\citenamefont {Sereno}(2004)}]{Sereno:2003nd}%
  \BibitemOpen
  \bibfield  {author} {\bibinfo {author} {\bibfnamefont {M.}~\bibnamefont
  {Sereno}},\ }\bibfield  {title} {\enquote {\bibinfo {title} {{Weak field
  limit of Reissner-Nordstrom black hole lensing}},}\ }\href {\doibase
  10.1103/PhysRevD.69.023002} {\bibfield  {journal} {\bibinfo  {journal} {Phys.
  Rev. D}\ }\textbf {\bibinfo {volume} {69}},\ \bibinfo {pages} {023002}
  (\bibinfo {year} {2004})},\ \Eprint {http://arxiv.org/abs/gr-qc/0310063}
  {arXiv:gr-qc/0310063 [gr-qc]}\BibitemShut {NoStop}%
\bibitem [{\citenamefont {Keeton}\ and\ \citenamefont
  {Petters}(2005)}]{Keeton:2005jd}%
  \BibitemOpen
  \bibfield  {author} {\bibinfo {author} {\bibfnamefont {C.~R.}\ \bibnamefont
  {Keeton}}\ and\ \bibinfo {author} {\bibfnamefont {A.~O.}\ \bibnamefont
  {Petters}},\ }\bibfield  {title} {\enquote {\bibinfo {title} {{Formalism for
  testing theories of gravity using lensing by compact objects: Static,
  spherically symmetric case}},}\ }\href {\doibase 10.1103/PhysRevD.72.104006}
  {\bibfield  {journal} {\bibinfo  {journal} {Phys. Rev. D}\ }\textbf {\bibinfo
  {volume} {72}},\ \bibinfo {pages} {104006} (\bibinfo {year} {2005})},\
  \Eprint {http://arxiv.org/abs/gr-qc/0511019} {arXiv:gr-qc/0511019
  [gr-qc]}\BibitemShut {NoStop}%
\bibitem [{\citenamefont {Keeton}\ and\ \citenamefont
  {Petters}(2006)}]{Keeton:2006sa}%
  \BibitemOpen
  \bibfield  {author} {\bibinfo {author} {\bibfnamefont {C.~R.}\ \bibnamefont
  {Keeton}}\ and\ \bibinfo {author} {\bibfnamefont {A.~O.}\ \bibnamefont
  {Petters}},\ }\bibfield  {title} {\enquote {\bibinfo {title} {{Formalism for
  testing theories of gravity using lensing by compact objects. II. Probing
  post-post-Newtonian metrics}},}\ }\href {\doibase 10.1103/PhysRevD.73.044024}
  {\bibfield  {journal} {\bibinfo  {journal} {Phys. Rev. D}\ }\textbf {\bibinfo
  {volume} {73}},\ \bibinfo {pages} {044024} (\bibinfo {year} {2006})},\
  \Eprint {http://arxiv.org/abs/gr-qc/0601053} {arXiv:gr-qc/0601053
  [gr-qc]}\BibitemShut {NoStop}%
\bibitem [{\citenamefont {Virbhadra}\ and\ \citenamefont
  {Ellis}(2002)}]{Virbhadra:2002ju}%
  \BibitemOpen
  \bibfield  {author} {\bibinfo {author} {\bibfnamefont {K.~S.}\ \bibnamefont
  {Virbhadra}}\ and\ \bibinfo {author} {\bibfnamefont {G.~F.~R.}\ \bibnamefont
  {Ellis}},\ }\bibfield  {title} {\enquote {\bibinfo {title} {{Gravitational
  lensing by naked singularities}},}\ }\href {\doibase
  10.1103/PhysRevD.65.103004} {\bibfield  {journal} {\bibinfo  {journal} {Phys.
  Rev. D}\ }\textbf {\bibinfo {volume} {65}},\ \bibinfo {pages} {103004}
  (\bibinfo {year} {2002})}\BibitemShut {NoStop}%
\bibitem [{\citenamefont {Virbhadra}\ and\ \citenamefont
  {Keeton}(2008)}]{Virbhadra:2007kw}%
  \BibitemOpen
  \bibfield  {author} {\bibinfo {author} {\bibfnamefont {K.~S.}\ \bibnamefont
  {Virbhadra}}\ and\ \bibinfo {author} {\bibfnamefont {C.~R.}\ \bibnamefont
  {Keeton}},\ }\bibfield  {title} {\enquote {\bibinfo {title} {{Time delay and
  magnification centroid due to gravitational lensing by black holes and naked
  singularities}},}\ }\href {\doibase 10.1103/PhysRevD.77.124014} {\bibfield
  {journal} {\bibinfo  {journal} {Phys. Rev. D}\ }\textbf {\bibinfo {volume}
  {77}},\ \bibinfo {pages} {124014} (\bibinfo {year} {2008})},\ \Eprint
  {http://arxiv.org/abs/0710.2333} {arXiv:0710.2333 [gr-qc]}\BibitemShut
  {NoStop}%
\bibitem [{\citenamefont {Gyulchev}\ and\ \citenamefont
  {Yazadjiev}(2008)}]{Gyulchev:2008ff}%
  \BibitemOpen
  \bibfield  {author} {\bibinfo {author} {\bibfnamefont {G.~N.}\ \bibnamefont
  {Gyulchev}}\ and\ \bibinfo {author} {\bibfnamefont {S.~S.}\ \bibnamefont
  {Yazadjiev}},\ }\bibfield  {title} {\enquote {\bibinfo {title}
  {{Gravitational lensing by rotating naked singularities}},}\ }\href {\doibase
  10.1103/PhysRevD.78.083004} {\bibfield  {journal} {\bibinfo  {journal} {Phys.
  Rev. D}\ }\textbf {\bibinfo {volume} {78}},\ \bibinfo {pages} {083004}
  (\bibinfo {year} {2008})},\ \Eprint {http://arxiv.org/abs/0806.3289}
  {arXiv:0806.3289 [gr-qc]}\BibitemShut {NoStop}%
\bibitem [{\citenamefont {Sahu}\ \emph {et~al.}(2012)\citenamefont {Sahu},
  \citenamefont {Patil}, \citenamefont {Narasimha},\ and\ \citenamefont
  {Joshi}}]{Sahu:2012er}%
  \BibitemOpen
  \bibfield  {author} {\bibinfo {author} {\bibfnamefont {S.}~\bibnamefont
  {Sahu}}, \bibinfo {author} {\bibfnamefont {M.}~\bibnamefont {Patil}},
  \bibinfo {author} {\bibfnamefont {D.}~\bibnamefont {Narasimha}}, and\
  \bibinfo {author} {\bibfnamefont {P.~S.}\ \bibnamefont {Joshi}},\ }\bibfield
  {title} {\enquote {\bibinfo {title} {{Can strong gravitational lensing
  distinguish naked singularities from black holes?}}}\ }\href {\doibase
  10.1103/PhysRevD.86.063010} {\bibfield  {journal} {\bibinfo  {journal} {Phys.
  Rev. D}\ }\textbf {\bibinfo {volume} {86}},\ \bibinfo {pages} {063010}
  (\bibinfo {year} {2012})},\ \Eprint {http://arxiv.org/abs/1206.3077}
  {arXiv:1206.3077 [gr-qc]}\BibitemShut {NoStop}%
\bibitem [{\citenamefont {Weinberg}(1972)}]{Weinberg:1972kfs}%
  \BibitemOpen
  \bibfield  {author} {\bibinfo {author} {\bibfnamefont {S.}~\bibnamefont
  {Weinberg}},\ }\href
  {http://www-spires.fnal.gov/spires/find/books/www?cl=QC6.W431} {\emph
  {\bibinfo {title} {{Gravitation and Cosmology: Principles and Applications of
  the General Theory of Relativity}}}}\ (\bibinfo  {publisher} {John Wiley and
  Sons},\ \bibinfo {address} {New York},\ \bibinfo {year} {1972})\BibitemShut
  {NoStop}%
\bibitem [{\citenamefont {Virbhadra}\ and\ \citenamefont
  {Ellis}(2000)}]{Virbhadra:1999nm}%
  \BibitemOpen
  \bibfield  {author} {\bibinfo {author} {\bibfnamefont {K.~S.}\ \bibnamefont
  {Virbhadra}}\ and\ \bibinfo {author} {\bibfnamefont {G.~F.~R.}\ \bibnamefont
  {Ellis}},\ }\bibfield  {title} {\enquote {\bibinfo {title} {{Schwarzschild
  black hole lensing}},}\ }\href {\doibase 10.1103/PhysRevD.62.084003}
  {\bibfield  {journal} {\bibinfo  {journal} {Phys. Rev. D}\ }\textbf {\bibinfo
  {volume} {62}},\ \bibinfo {pages} {084003} (\bibinfo {year} {2000})},\
  \Eprint {http://arxiv.org/abs/astro-ph/9904193} {arXiv:astro-ph/9904193
  [astro-ph]}\BibitemShut {NoStop}%
\bibitem [{\citenamefont {Brans}(1962)}]{Brans:1962zz}%
  \BibitemOpen
  \bibfield  {author} {\bibinfo {author} {\bibfnamefont {C.~H.}\ \bibnamefont
  {Brans}},\ }\bibfield  {title} {\enquote {\bibinfo {title} {{Mach's principle
  and a relativistic theory of gravitation. II}},}\ }\href {\doibase
  10.1103/PhysRev.125.2194} {\bibfield  {journal} {\bibinfo  {journal} {Phys.
  Rev.}\ }\textbf {\bibinfo {volume} {125}},\ \bibinfo {pages} {2194--2201}
  (\bibinfo {year} {1962})}\BibitemShut {NoStop}%
\bibitem [{\citenamefont {Bhadra}\ and\ \citenamefont
  {Sarkar}(2005)}]{Bhadra:2005mc}%
  \BibitemOpen
  \bibfield  {author} {\bibinfo {author} {\bibfnamefont {A.}~\bibnamefont
  {Bhadra}}\ and\ \bibinfo {author} {\bibfnamefont {K.}~\bibnamefont
  {Sarkar}},\ }\bibfield  {title} {\enquote {\bibinfo {title} {{On static
  spherically symmetric solutions of the vacuum Brans-Dicke theory}},}\ }\href
  {\doibase 10.1007/s10714-005-0181-1} {\bibfield  {journal} {\bibinfo
  {journal} {Gen. Rel. Grav.}\ }\textbf {\bibinfo {volume} {37}},\ \bibinfo
  {pages} {2189--2199} (\bibinfo {year} {2005})},\ \Eprint
  {http://arxiv.org/abs/gr-qc/0505141} {arXiv:gr-qc/0505141
  [gr-qc]}\BibitemShut {NoStop}%
\end{thebibliography}
\end{document}